\documentclass[letter]{aa}

\usepackage{natbib}
\usepackage{graphicx}
\usepackage{txfonts}
\usepackage{color}

\newcommand{\msun}{\mbox{$\,{\rm M}_\odot$}}

\begin{document}

\title{VLT/SPHERE deep insight of NGC\,3603's core\thanks{Based on data collected at the European Southern Observatory, Chile (Guaranteed Time Observation 095.D-0309(A) and 095.D-0309(E))}: \\ Segregation or confusion?}
\author{
        Z. Khorrami\inst{1}
        \and
        T. Lanz \inst{1}
        \and
        F. Vakili \inst{1}
        \and
        E. Lagadec \inst{1}
        \and
        M. Langlois \inst{2,}\inst{3}
        \and
        W. Brandner \inst{4}
        \and
        O. Chesneau \inst{1}
        \and
        M. R. Meyer \inst{5}
        \and
        M. Carbillet \inst{1}
      \and
        L. Abe \inst{1}
        \and 
        D. Mouillet \inst{6}
       \and
        JL. Beuzit  \inst{6}
      \and
        A. Boccaletti \inst{7}
        \and
        C. Perrot \inst{7}
        \and
        C. Thalmann \inst{5}
        \and
        H.-M. Schmid \inst{5}
        \and
        A. Pavlov \inst{4}
        \and
        A. Costille \inst{3}
        \and
        K. Dohlen\inst{3}
        \and
        D. Le Mignant \inst{3}
        \and
        C. Petit \inst{8}
        \and
        J.F. Sauvage \inst{8}
}

\institute{
        Laboratoire Lagrange, Universit\'{e} C\^{o}te d'Azur, Observatoire de la C\^{o}te d'Azur, CNRS, F-06304 Nice, France.\\
 \email{zeinab.khorrami@oca.eu}
        \and
        Univ Lyon, Univ Lyon1, Ens de Lyon, CNRS, CRAL UMR5574, F-69230, Saint-Genis-Laval, France
        \and
        Aix Marseille Université, CNRS, LAM - Laboratoire d’Astrophysique de Marseille, UMR 7326, 13388, Marseille, France
        \and
        Max-Planck-Institut fur Astronomie, Konigstuhl 17, 69117 Heidelberg, Germany
       \and
        Institute for Astronomy, ETH Zurich, Wolfgang-Pauli-Strasse 27, CH-8093 Zurich, Switzerland
      \and
      Université Grenoble Alpes, CNRS, IPAG, 38000 Grenoble, France
       \and
        LESIA, Observatoire de Paris, CNRS, Université Paris 7, Université Paris 6, 5 place Jules Janssen, 92190 Meudon, France
       \and
       ONERA - Optics Department, 29 avenue de la Division Leclerc, F-92322 Chatillon Cedex, France
}

\date{}

\abstract{
We present new near-infrared photometric measurements of the core of the young massive cluster NGC\,3603 obtained with extreme adaptive optics. The data were obtained with the SPHERE instrument mounted on ESO's Very Large Telescope, and cover three fields in the core of this cluster. We applied a correction for the effect of extinction to our data obtained in the J and K broadband filters and estimated the mass of detected sources inside the field of view of SPHERE/IRDIS, which is 13.5"$\times$13.5". 
We derived the mass function (MF) slope for each spectral band and field. The MF slope in the core is unusual compared to previous results based on Hubble space telescope (HST) and very large telescope (VLT) observations. 
The average slope in the core is estimated as $-1.06\pm0.26$ for the main sequence stars with 3.5\msun < M < 120\msun.
Thanks to the SPHERE extreme adaptive optics, 814 low-mass stars were detected to estimate the MF slope for the pre-main sequence stars with 0.6\msun < M < 3.5\msun, $\Gamma = -0.54 \pm 0.11$ in the K-band images in two fields in the core of the cluster. 
For the first time, we derive the mass function of the very core of the NGC\,3603 young cluster for masses in the range 0.6 - 120 \msun.
 Previous studies were either limited by crowding, lack of dynamic range, or a combination of both.
}
   \maketitle

\section{Introduction} 

Among Galactic spiral arm clusters, the NGC\,3603 young cluster, located in its namesake giant HII region \citep{kennicutt84}, is the most compact and youngest cluster with an age of 1 Myr \citep{{Brandl99},{Stolte04},{Sung04}} and a central density of $6\times10^4$ \msun\, $\rm{pc}^{-3}$ \citep{Harayama08}. 
The cluster is known to contain three massive and luminous central stars with spectral types as early as O2V \citep{{walborn2002}, {moffat2004}}, and up to 50 O-type stars in total \citep{Drissen95}.
The most massive stars exhibit both characteristics of WN6 stars and Balmer absorption lines \citep{Drissen95}, suggesting that they are actually core hydrogen burning rather than evolved stars \citep{{Conti95},{deKoter97}}. Two of these three Wolf-Rayet (WR) stars are very close binaries \citep{Schnurr08}.
The total mass of the cluster is estimated as $10^4 \msun$ \citep{Harayama08}, with an upper limit to the dynamical mass of $17600 \pm 3800$\msun \citep{Rochau10}. 
NGC\,3603 provides a unique opportunity to study the formation of massive stars embedded in clusters at their early stages. Studying such clusters is not generally straightforward owing to the limited angular resolution of telescopes in addition to uncertainties from existing models \citep{Ascenso2009}. 
Besides, extinction from the Galactic plane and  the birth place of massive stars, which is immersed in dust and gas, play an important role.

All these combined effects can introduce an observational bias that hampers differentiating models of massive star and cluster formation: i.e., singular collapse of a rotating molecular cloud core with subsequent fragmentation in a flattened disk or competitive accretion, or, for example, external triggering by cloud-cloud collision (ref., e.g., Johnston et al. 2014).
To test these models, high angular resolution observation in the infrared are the best strategy as they minimize the effects of source confusion and spatial extinction variations.

In this context, the extreme adaptive optics (XAO) of the new instrument SPHERE \citep{sphere} on the VLT, enabled us to observe deeper in the core of NGC\,3603 in the near-infrared J and K bands to better probe the massive star cluster at a high resolution in the range of 20-40 mas resolution, which is comparable to the HST in the visible.
\section{Data and photometry}\label{sec:dataphot}
We obtained data via Guaranteed Time Observation (GTO) runs to image NGC\,3603 using the dual mode of IRDIS \citep{maud14}, enabling simultaneous observations in two spectral bands on the VLT. 
The observations were performed in two epochs in 2015 (March and June), with high dynamic and spatial resolution imaging
of three regions, each with a field of view (FoV) of $13.5"\times13.5"$, one centered on the core of the cluster (F0) and two regions (F1 and F2) to cover the radial extent of the cluster (Figure \ref{fig:fov} Top).
To facilitate homogeneous photometric calibrations, F1 and F2 partially overlap with F0.
The data consist of 400, 300, and 320 frames of 0.8s exposures in the IRDIS broadband K filter (IRDIS/BB-K)  for F0, F1, and F2, respectively, and 400, 150, and 160 frames of 4.0, 2.0, and 2.0s exposures in IRDIS broadband J (IRDIS/BB-J), respectively, during the two observing epochs. Neutral density (ND) filters were used for the IRDIS/BB-J data to avoid saturating the brightest stars.
The average airmass during these observations was 1.25-1.26. A log of the observations is presented in Table \ref{table:expo2}. 

We used the SPHERE pipeline package\footnote{\url{http://www.mpia.de/SPHERE/sphere-web/nightly_builds-page.html}}, for correcting dark, flat, distortion, and bad pixels. As SPHERE filters in BB-J and BB-K are similar to ESO's Nasmyth Adaptive Optics System Near-Infrared Imager and Spectrograph (NACO), we corrected the photometric zero-points of SPHERE by comparing them to the magnitudes of spectroscopically known stars (preferentially isolated sources) in NACO \citep{Harayama08} J and K filters. 

For the photometry, we used the STARFINDER package implemented in IDL \citep{starfinder} to derive the local point spread function (PSF) to detect stellar objects, while estimating instrumental magnitudes, i.e., before the photometric zero-point corrections. 
For this, each field is divided into subregions  to extract the empirical local PSFs from isolated sources in the image itself. Local PSFs are then used to extract the flux of the sources  to compensate for the local distortion effect.
This is particularly suitable for AO-assisted imaging data where one can face locally distorted PSFs that hamper photometric measurements along different parts of the IRDIS FoV.

Consequently, 410 (290), 149 (364), and 445 (682) sources were detected in the J and K bands in F0, F1, and F2, respectively, by limiting them to a minimum correlation of 0.8 with the PSF. We also put a threshold limit of one standard deviation of sky brightness. Table \ref{table:slope-irdis} gives the details of the total number of detected sources in each field for a given filter. 

The high Strehl ratio, and the resulting high dynamic range close to bright stars, enabled us to detect stellar sources that are 10.6 and 9.8 magnitudes fainter than the brightest sources in J and K bands, respectively. Sources with K magnitude of 18.8 and J magnitude of 18.9 could be detected in the core of NGC\,3603. 
Our test experiments for completeness correction (500 artificial star per flux) suggest that we should reach a completeness level $>=80\%$ for stars brighter than 17.5 mag in F1 and F2 in both J and K bands.
The quality (signal-to-noise Ratio) of data in F0 in K band (in March 2015) was not as good as in F1 and F2 in the second run (June 2015), thus the dynamic range is lower. In F0, stars brighter than 15.3 in K band and 16.5 in J band have a completeness level of $>=80\%$.
Table \ref{table:expo2} gives the faintest magnitudes obtained in the different fields F0, F1, and F2 by SPHERE within the exposure time limits of the run. 
\begin{table}
\caption{Exposure time log and faintest stars (SNR > 4.2) of VLT/SPHERE observations.
$\Delta_{mag}$ is the difference between the maximum and minimum magnitudes in F0, F1, and F2 fields.}
\begin{tabular}{ c c c  c c c c}  
Field&Single/Total&$\lambda_{cen}$&$\Delta_\lambda$&$mag_{max}$&mass&$\Delta_{ mag}$\\
(Filter)&Exposure[s] & [nm]  & [nm]    &    & [M$_{\odot}]$ &  \\
\hline
F0 (J)&4.0/1600   &1245&240&18.7&0.66&10.6\\
F0 (K)&0.83/335.0&2182&300&16.4&1.08&9.5\\
F1 (J)&2.0/300.0  &1245&240&18.9&0.57&9.0\\
F1 (K)&0.83/251.3&2182&300&18.1&0.29&9.8\\
F2 (J)&2.0/320.0  &1245&240&18.9&0.58&9.3\\
F2 (K)&0.83/269.0&2182&300&18.8&0.14&9.6\\
\end{tabular} 
\label{table:expo2}  
\end{table}
\section{Extinction and CMD}\label{redcor}
\begin{figure}
\centering
\includegraphics[width=7.cm]{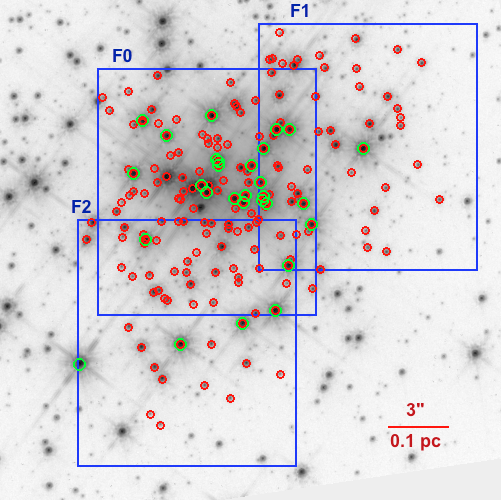}\\
\includegraphics[width=7.3cm]{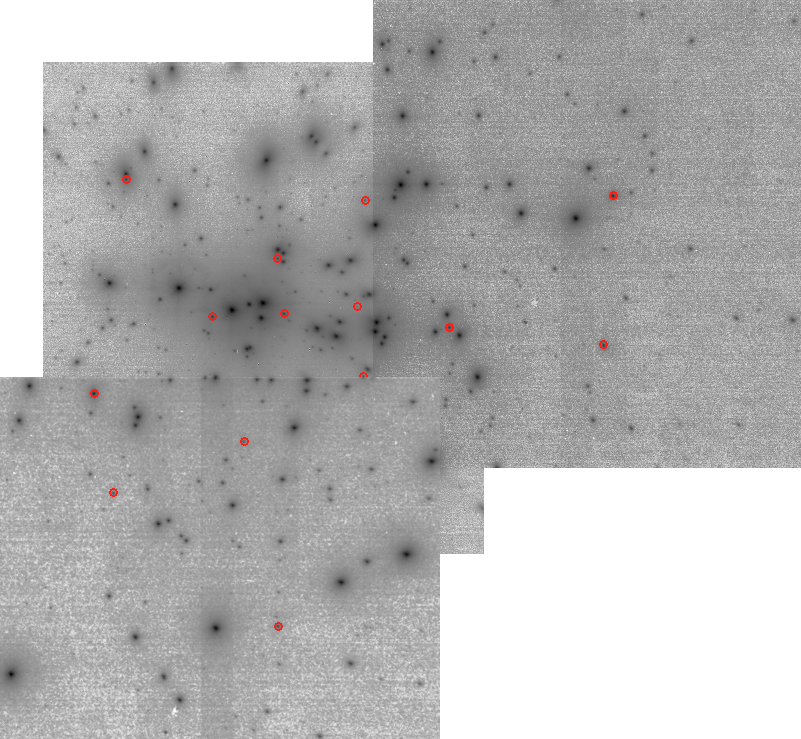}
\caption{Top: HST/WFPC2-PC/F814W core of NGC\,3603: blue squares depict the three fields observed with SPHERE-IRDIS. Green circles refer to the known O-type stars from \cite{Harayama08}.  Red circles show the stars in \cite{Harayama08} catalog, which we used for calibrating zero-points in different fields. There are 113, 45, and 51 stars common between \cite{Harayama08} in the SPHERE F0, F1, and F2 fields.
Bottom: sources with $E(J-K) > 1.8$ in F0, F1, and F2 shown in red circles. The image is a combination of all three fields in IRDIS/B-J.}
\label{fig:fov}
\end{figure}

In order to correct for extinction, 31 O stars on or close to the main sequence (class V) were selected from \cite{Harayama08}. These stars are encircled in green in the top panel of Figure \ref{fig:fov}.  
To estimate their intrinsic magnitude, their $\rm{T}_{\rm{eff}}$ were estimated from Table 4 of \cite{martins} and their $\log{\rm{g}}$ as a function of age (1.5 Myr) according to the PARSEC \footnote{http://stev.oapd.inaf.it/cgi-bin/cmd} stellar evolution models \citep{Bressan12}, adopting Galactic metallicity. We assumed a distance of 6 kpc (Section \ref{sec:mf}) for these O stars.

We derived the color excess for selected O stars in the two IRDIS-BB J and K filters at the distance of 6 kpc.
We adopted the maximum weighted value for E(J-K), which is 0.76 (A$_{\rm{V}} = 4.5$). This value results in A$_{\rm{J}}$ and A$_{\rm{K}}$ as 1.269 and 0.504, respectively, (from \cite{rieke}).
\begin{figure}
\includegraphics[trim=10 0 10 6,clip,width=8.5cm]{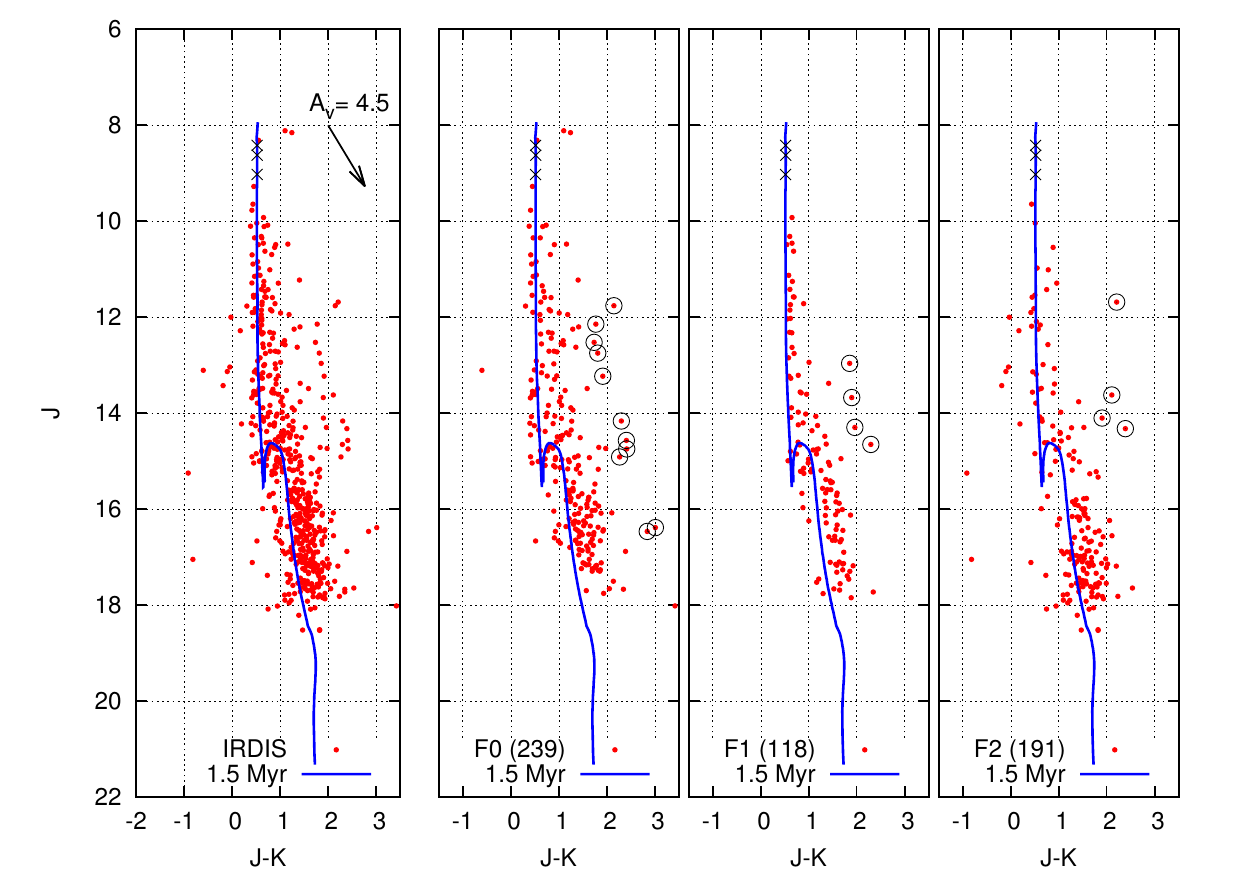}
\caption{CMD of the core of NGC\,3603 in IRDIS J and K band for the whole FoV (left) followed to the right for the three fields F0, F1, and F2. Black circles show the K-excess stars. Three black crosses represent stellar models with initial masses of 100, 120, and 150 \msun. The black arrow signifies the effect of extinction, A$_{\rm{V}} = 4.5$.}
\label{fig:cmdirdis}
\end{figure}
We found 239, 118, and 191 sources detected in both J and K data in F0, F1, and F2, respectively. The color magnitude diagrams (CMDs) for these different fields are shown in Figure \ref{fig:cmdirdis}.
\section{Mass functions}\label{sec:mf}
We used the stellar evolution tracks from PARSEC mentioned above to estimate stellar masses at the age of 1.5 Myr. The distance of the cluster was taken as 6 kpc, which fits well with the observed CMD, isochrone, and extinction, and is also in good agreement with \cite{Stolte04}, \cite{Depree99}, and \cite{Harayama08}.

\begin{figure}
\includegraphics[trim=11 35 9 35,clip,width=8.6cm]{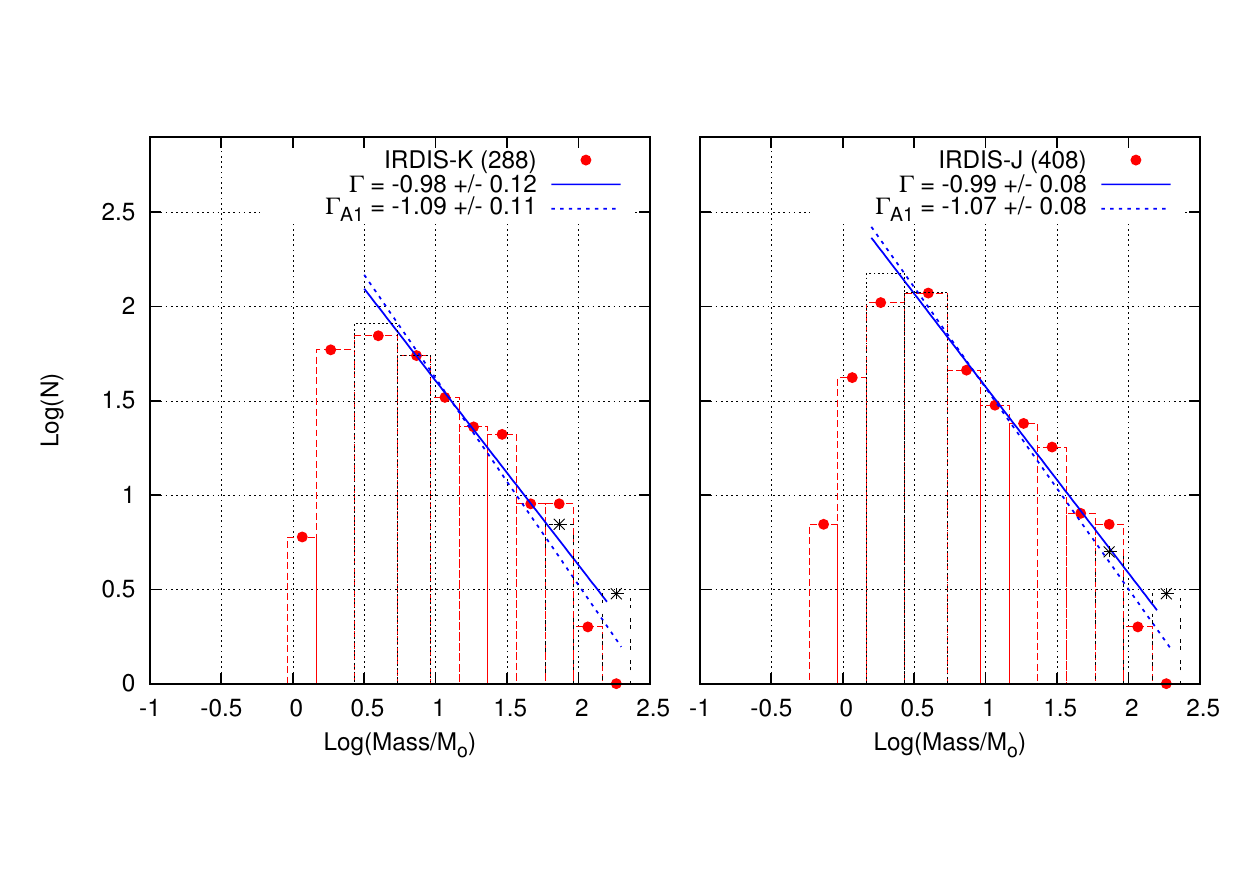}
\caption{Mass functions derived for IRDIS data in BB-J (right) and BB-K (left) in F0 (shown in Figure \ref{fig:fov} Top). Last three red bins represent the MF considering the A1 and C as a binaries and B as a single source. The last three black stars represent the MF if the WR stars are considered single objects, which is the case for the photometry analysis.}
\label{fig:mfirdis0}
\end{figure}

\begin{figure}
\includegraphics[trim=11 35 9 35,clip,width=8.6cm]{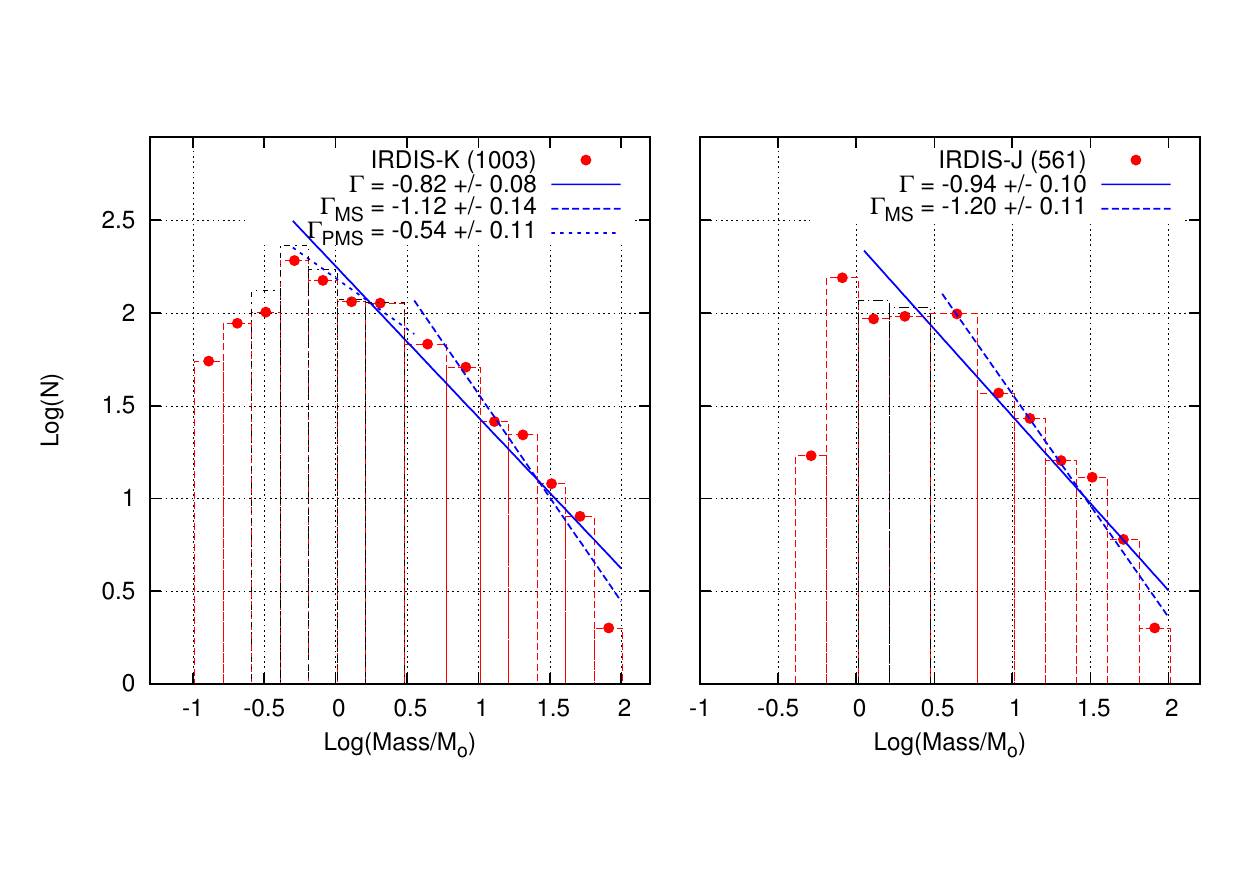}
\caption{Mass functions derived for IRDIS data in BB-J (right) and BB-K (left) in F1 and F2 together. F1 and F2 are shown in Figure \ref{fig:fov} Top.}
\label{fig:mfirdis12}
\end{figure}

Using grids of stellar evolutionary tracks and extinction for each filter, we can estimate the stellar masses separately from the photometry in each filter. The slope of the mass function $\Gamma$ can be estimated from Eq. \ref {eq:mf}, where M is the stellar mass and N is the number of stars in the logarithmic mass interval $\log_{10}(M)$ to $\log_{10}(M)+0.2\log_{10}(M)$.
We used a double size bin at the pre-main sequence (PMS) and main sequence (MS) transition, around 4 \msun, to smooth out the degeneracy (same as \cite{Harayama08}).
We used an implementation of the nonlinear least-squares (NLLS) Marquardt-Levenberg algorithm to fit the MF,
\begin{equation}  
\log(\rm{N})=\Gamma~\log(\frac{M}{~M_{\odot}})+cst.
\label{eq:mf}
\end{equation} 
Mass functions for the two IRDIS BB-J and BB-K filters are shown in Figure \ref{fig:mfirdis0} for the core (F0) and in Figure \ref{fig:mfirdis12} for F1 and F2.  One can compare the MF in the very core of the cluster (F0) with the next radial bin (F1 and F2) to check the radial variation of MF.
The two latter were observed with similar exposure times and very close conditions as both fields were recorded within one hour slot of a SPHERE/VLT run on  2015-06-07. Also, minimum and maximum magnitudes in both fields are very similar especially in the K band. All these conditions result in similar mass range of detected sources in J and K. 
Thus we could plot the MF for F1 and F2 together, where 586 sources are detected with masses less than 1 $M_{\odot}$ (fainter than 16.5 magnitude) in the K band.
 
In F0, we were able to reach 0.66 $M_{\odot}$ (J=18.7) in the J band and 1.08 $M_{\odot}$ (K=16.4) in the K band.
Figure \ref{fig:mfirdis0} depicts the MF in F0. 
Three WR stars (A1, B, C) are located in this region where two of them (A1 and C) were identified as spectroscopic binaries by \cite{Schnurr08}. The MF can be treated in three possible ways (Table \ref{table:slope-irdis}): 
1) considering all WR stars: two as binaries with masses estimated from \cite{Schnurr08} and one (B) as a single star ({\it{All}}); 
2) considering just two WR stars as two binary systems ({\it{All-B}}); and 
3) excluding these three WR stars ({\it{All-WRs}}).

Figure \ref{fig:mfirdis12} depicts the MF for the next radial bin from the core of NGC3603 (F1 and F2). The mass range covered in K band starts from 0.14 \msun, ending at 69 \msun. More than 800 sources with a mass smaller than 4 \msun ~are detected. 

The change of the MF slope for the MS and PMS stars occurs around 4 \msun.
We also fitted a separate line on MF (dash-dotted line in Figure \ref{fig:mfirdis12}) for the low-mass PMS stars. 
The mass function in the low-mass part is corrected for the number of detected sources above an incompleteness level of 80\% (black bins in the low-mass part in Figure \ref{fig:mfirdis0} and \ref{fig:mfirdis12}).

Table \ref{table:slope-irdis} lists the derived MF slopes in F0, F1, and F2 regions and derived slopes for MS/PMS stars. MS is a common mass range in J and K and in F0 and F1+F2 with an incompleteness level of 100\%.
The MF slopes are consistent in the J band and K band and also in the different regions, for the main sequence stars and for the whole mass range. 

The MF slopes for the whole mass range is flatter than the main sequence part.
The MF slopes even for the massive stars (main sequence) are flatter than Salpeter, $\Gamma_{Salpeter}=-1.35$ \citep{salpeter} and Kroupa, $\Gamma_{Kroupa}=-1.3$ \citep{kroupa2001}. The average value agrees with those found in the outer regions of NGC3603 according to previous works.
For pre-main sequence stars, the MF slope is flatter than for the whole mass range and for the main sequence.

If we assume that binary properties like binary fraction and mass-ratio distribution do not change strongly with the mass of the primary stars, then the deduced mass function slope should be very similar to the mass function slope of the primary stars.

We could detect 11, 4, and 4 K-excess sources with an E(J-K) that is higher than 1.8 (for MS) and 2.0 (for PMS), in F0, F1, and F2, respectively, corresponding to 14 sources in total (black circles in Figure \ref{fig:cmdirdis} and red circles in the bottom panels of Figure \ref{fig:fov}). For these stars, the mass estimated in K is higher than in J because of their K excess. Twelve of these stars are on the sequence parallel to the MS ($M > 4$ \msun).  
In this case, the MF slopes for the main sequence part in the J band should be more reliable than in the K band.
 
\begin{table}
\caption{Number of detected stars ($N_J$ and $N_K$ in J and K bands) and MF slopes ($\Gamma_{J}$ and $\Gamma_{K}$ in J and K bands) using Equation \ref{eq:mf}, at 1.5 Myr in three F0, F1, and F2 fields of NGC\,3603 from SPHERE/IRDIS.} 
\begin{tabular}{c c c c c} 
F0&$N_K$&$\Gamma_{K}$&$N_J$&$\Gamma_{J}$\\
\hline 
{\it{All}}&288&$-1.09\pm0.11$&408&$-1.07\pm0.08$\\
{\it{All-B}}&287&$-0.98\pm0.12$&407&$-0.99\pm0.08$\\
{\it{All-WRs}}&283&$-0.85\pm0.06$&403&$-0.98\pm0.09$\\
\hline
F1+F2&$N_K$&$\Gamma_{K}$&$N_J$&$\Gamma_{J}$\\
\hline
Total&1003&$-0.82\pm0.08$&561&$-0.94\pm0.10$\\
MS&189&$-1.12\pm0.14$&200&$-1.20\pm0.11$\\
PMS&814&$-0.54\pm0.11$&361&-\\
\end{tabular} 
\label{table:slope-irdis}  
\end{table}
\section{Discussion and conclusion}
NGC\,3603 has been observed with various instruments and the slope of its mass function has been calculated in previous works (Table \ref{table:slopeshistory}). These works reach conclusions on the general trend of decreasing MF slopes in the core as the signature of mass segregation. 
The slopes of the MF in different filters (Table \ref{table:slope-irdis}) in the core of NGC\,3603 is steeper than the previous works and does not show a radial dependence in the observed fields. The slope value for the main sequence stars and for the whole mass range is consistent in all observed regions of the core of the cluster.

Shape of MF at the massive end, can be used as an observational test that may be able to settle the question of which mechanism (accretion or collision) is a dominant route for the formation of the most massive stars \citep{krumholz}. 
According to the accretion models, as the massive stars form by the same accretion processes that produce low-mass stars (normal star formation), the high end of the stellar mass function should be continuous and does not depend radically on the environment \citep{krumholz}. 
On the other hand, collisional formation predicts a large gap in the stellar MF, separating the bulk of the accretion-formed stellar population from the few collisionally formed stars \citep{{Baumgardt11},{Moeckel11}}. This feature should only appear in the most massive and densest clusters. Figure \ref{fig:mfirdis0} shows this signature in the core (F0), but we know that the last bin corresponds to the three WR stars in which two of them have been found to be multiple objects and not single stars \citep{Schnurr08}. Therefore, collisional formation of very massive
objects seems unlikely at least for the NGC\,3603 cluster.
Accretion models also predict that massive stars are likely to have low-mass and high-mass companions \citep{{kratter2006},{kratter2008},{kratter2010},{krumholz2012}}, but the collisionally formed stars lack low-mass companions, which provokes segregation.

This study shows no signature of mass segregation in the core of NGC\,3603, first, because   the MF slope in its very core is not flatter than the next radial bin. Second, both slopes are similar to the MF values found in previous works for the outer regions (references in Table \ref{table:slopeshistory}).
Therefore, it appears that nonsegregated clusters with a smooth MF agree better with accretion models for massive star formation. 
Our SPHERE results demonstrate that, by improved photometric dynamic range and spatial resolution from XAO, we can overcome the effect of confusion that in the past has led to the conclusion of observational segregation (see also \cite{Ascenso2009}) as far as NGC\,3603 is concerned.

\begin{table}
\caption{Slopes of the mass function derived for NGC3603 in earlier works.} 
\begin{tabular}{ c c c } 
 $\Gamma$ & condition & reference \\
\hline
$-0.5 \pm 0.1$       &$r < 6"$, (1.6-100)\msun&\cite{Sung04}\\
-0.31                      &$0-5"$  , (0.4-20)\msun &\cite{Harayama08} \\
-0.26                      &$0-5"$  , (6.3-100)\msun&\cite{pang} \\
\hline
\end{tabular} 
\label{table:slopeshistory}  
\end{table}

{\it{Acknowledgements.}}
{\small{ZK is supported by the Erasmus Mundus Joint Doctorate Program by Grant Number 2012-1710 from the EACEA of the European Commission. We warmly thank Alain Chelli for useful discussions. We thank the anonymous referee for constructive comments, which  improved the paper.}}
\bibliographystyle{aa}
\bibliography{NGC3603final}
\end{document}